\documentclass[useAMS,usenatbib]{mn2e}

\usepackage{graphicx}

\title[The aerosols thermal emissivity in the NIR]{Near-infrared thermal emissivity from ground based atmospheric dust measurements at ORM\thanks{Based on data collected at TNG.}}
\author[G. Lombardi et al.]{\vspace{1.7mm}
G. Lombardi$^{1}$\thanks{E-mail: glombard@eso.org}, 
V. Zitelli$^{2}$, S. Ortolani$^{3}$, J. Melnick$^{1}$, A. Ghedina$^{4}$, A. Garcia$^{4}$, \\ 
\vspace{2.5mm}{\LARGE\rm E. Molinari$^{4}$ and C. Gatica$^{5}$} \\
$^{1}$European Southern Observatory, Casilla 19001, Santiago 19, Chile\\
$^{2}$INAF - Bologna Astronomical Observatory, via Ranzani 1, I-40127 Bologna, Italy\\
$^{3}$Department of Astronomy, University of Padova, vicolo dell'Osservatorio 2, I-35122 Padova, Italy\\
$^{4}$Fundaci\'{o}n Galileo Galilei and Telescopio Nazionale Galileo, Rambla Jos\`{e} Ana Fern\'{a}ndez P\'{e}rez 7, E-38712 Bre\~{n}a Baja, Tenerife, Spain\\
$^{5}$Department of Astronomy and Astrophysics, Pontificia Universidad Catolica, Av. Vicu\~{n}a Mackenna 4860, Santiago de Chile, Chile
}

\begin{document}
\date{Accepted 2011 May 29. Received 2011 May 29; in original form 2010 November 9}

\pagerange{\pageref{firstpage}--\pageref{lastpage}} \pubyear{2010}

\maketitle

\label{firstpage}
\setcounter {figure} {0}
\begin{abstract}
We present an analysis of the atmospheric content of aerosols measured at Observatorio del Roque de los Muchachos (ORM; Canary Islands). Using a laser diode particle counter located at the Telescopio Nazionale Galileo (TNG) we have detected particles of 0.3, 0.5, 1.0, 3.0, 5.0 and 10.0 $\mu$m size.\\
The seasonal behavior of the dust content in the atmosphere is calculated. The Spring has been found to be dustier than the Summer, but dusty conditions may also occur in Winter.\\
A method to estimate the contribution of the aerosols emissivity to the sky brightness in the near-infrared (NIR) is presented. The contribution of dust emission to the sky background in the NIR has been found to be negligible comparable to the airglow, with a maximum contribution of about 8-10\% in the $K_{S}$ band in the dusty days.
\end{abstract}


\section{Introduction}
Superb observing conditions are crucial in order to obtain the best scientific output form the next generation of ground based telescopes, and this requires monitoring all relevant parameters that may affect observations. Due to their potentially detrimental impact on astronomical observations, atmospheric aerosols are among the most important parameters to be monitored in modern site testing campaigns. The performance and the safety of telescopes depend on the presence of atmospheric dust, which may deposit on  mirrors, increase atmospheric extinction, and emits in the infrared (IR) bands thus increasing the sky brightness.

Several studies about the atmospheric radiative effects of mineral dust exist.
Gelado et al. \cite{gelado03} analysed the mean aerosols content at Gran Canaria (Canary Islands) in periods 1997-1998 and 2002-2003. A mean 
grain size of 0.6-4.9 $\mu$m and  a large annual variability in both density and size distribution have been found.

Cuevas\&Baldasano \cite{cuevas09} provided information on the incidence of the dust loaded African air masses at Observatorio del Teide in Tenerife (Canary Islands) and
at Observatorio del Roque de Los Muchachos (ORM) in La Palma (Canary Islands). The analysis has been made using the high quality observations performed by the Iza\~na atmospheric observatory. The in situ measurements for the 2002-2008 period show that the air is only partially affected by some dust
 loaded by African air mass intrusion. A significant PM10 concentration has been found only above the 80 percentile (26 $\mu$g m$^{-3}$) at Iza\~na.
 
\begin{table*}
   \begin{center}
    \caption[]{Main characteristics of the dust monitors.}
    \begin{tabular}{r | l | l}
    \hline
    & ABACUS TM301 & LASAIRII 310B\\
    \hline
            input flow rate  & 0.1 Cubic Foot Minute            & 1.0 Cubic Foot Minute\\
            size channels    & 0.3, 0.5, 1.0, 5.0 $\mu$m        & 0.3, 0.5, 1.0, 3.0, 5.0, 10.0 $\mu$m\\
            light source     & laser diode ($\lambda = 780$ nm) & laser diode ($\lambda = 780$ nm)\\
            sample rate      & 1 data per minute                & 1 data per 2 hours\\
            output           & RS-232                           & Ethernet\\
            running time     & August 2001 $-$ December 2006    & March 2007 $-$ today (January 2010 for this paper)\\
      \hline
      \end{tabular}
      \label{counters}
      \end{center}
 \end{table*}
   
The first analyses of the optical properties of dust and their impact on astronomy were done by Murdin \cite{murdin85}, Stickland et al. \cite{stickland87}, Guerrero et al. \cite{guerrero98}, and Jimenez et al. \cite{jimenez98}. More recently, Lombardi et al. \cite{lombardi08} (hereafter Paper I)  analyzed 5+ years of time-series data spanning the period August 2001 to December 2006 obtained with the Telescopio Nazionale Galileo (TNG) dust counter, and showed that dust particles increase the extinction in the $B$, $V$ and $I$ bands at ORM.

In the present paper we extend the analysis of Paper I by adding almost 3 years of new data (March 2007 to January 2010) to cover a total period of almost 8 years. In the first part of the paper we calculate the seasonal trend of the atmospheric aerosol content on yearly and monthly basis. In the second part of the paper we present a method to estimate the thermal emissivity of the dust in the near-infrared (NIR).

\section{Dust monitors}\label{monitors}
Since 2001 the TNG site monitoring group has used two different particle counters made by Particle Measuring System to monitor the dust content of the atmosphere around the telescope. The Abacus TM301 
measured between August 2001 and December 2006, and the LasairII 310B is in operation since March 2007.

The data from the Abacus TM301 were extensively analyzed and presented in Paper I. The LasairII 310B provides a significant improvement over its predecessor, in fact it measures 6 different  particle sizes (0.3, 0.5, 1.0, 3.0, 5.0 and 10.0 $\mu$m) instead of only 4 measured by  the Abacus TM301.

Table \ref{counters} summarizes the basic instrumental characteristics of these counters that use a laser scattering technique for environmental ambient air analysis. Both are compact and portable devices designed to measure the air-purity of closed environments such as clean rooms  (Porceddu et al.\citealt{porceddu02}; Ghedina et al.\citealt{ghedina04}; Paper I). The light scattered by the dust particles is  converted to voltage pulses of amplitude proportional to particle size and frequency proportional to particle density. A long silicon pipe through the enclosure feeds the pump with external air at 13 m above the ground, that correspond to the level of the TNG primary mirror.

\begin{table*}
    \begin{center}
      \caption[]{Dust content in [N m$^{-3}$] at ORM in wintertime, summertime, and in the entire annual cycle for cases \textbf{(1)}, \textbf{(2)} and 		\textbf{(3)} (see text in Section \ref{season_sec}).}
        \begin{tabular}{r c c c c c c c c c c c}
        \hline
        \textbf{CASE (1)} & & & & BACKGROUND      & & & & DUSTY & & & statistics\\
               & & & Wintertime & Summertime & Annual & & & Annual & & & running time\\
        \hline
        0.3 $\mu$m & & & $1.3 \times 10^{6}$ & $4.4 \times 10^{6}$ & $3.0 \times 10^{6}$ & & & $1.8 \times 10^{7}$ & & & Aug. 2001 $-$ Dec. 2006\\
        0.5 $\mu$m & & & $1.2 \times 10^{5}$ & $3.8 \times 10^{5}$ & $2.5 \times 10^{5}$ & & & $6.1 \times 10^{6}$ & & & "\\
        1.0 $\mu$m & & & $0.5 \times 10^{5}$ & $1.5 \times 10^{5}$ & $1.0 \times 10^{5}$ & & & $4.1 \times 10^{6}$ & & & "\\
        5.0 $\mu$m & & & $0.7 \times 10^{3}$ & $1.5 \times 10^{3}$ & $1.1 \times 10^{3}$ & & & $1.6 \times 10^{5}$ & & & "\\
        \hline \vspace{-0.2cm}
        & & & & & & & &\\
         \hline
        \textbf{CASE (2)} & & & & BACKGROUND      & & & & DUSTY & & & statistics\\
               & & & Wintertime & Summertime & Annual & & & Annual & & & running time\\
        \hline
        0.3 $\mu$m & & & $1.1 \times 10^{6}$ & $3.7 \times 10^{6}$ & $2.3 \times 10^{6}$ & & & $3.0 \times 10^{7}$ & & & Mar. 2007 $-$ Jan. 2010\\
        0.5 $\mu$m & & & $1.2 \times 10^{5}$ & $3.7 \times 10^{5}$ & $2.5 \times 10^{5}$ & & & $4.1 \times 10^{6}$ & & & "\\
        1.0 $\mu$m & & & $0.5 \times 10^{5}$ & $1.0 \times 10^{5}$ & $0.5 \times 10^{5}$ & & & $2.6 \times 10^{6}$ & & & "\\
        3.0 $\mu$m & & & $2.7 \times 10^{3}$ & $5.2 \times 10^{3}$ & $3.9 \times 10^{3}$ & & & $5.1 \times 10^{5}$ & & & "\\
        5.0 $\mu$m & & & $1.2 \times 10^{3}$ & $1.6 \times 10^{3}$ & $1.5 \times 10^{3}$ & & & $1.6 \times 10^{5}$ & & & "\\
       10.0 $\mu$m & & & $1.2 \times 10^{2}$ & $0.7 \times 10^{2}$ & $1.0 \times 10^{2}$ & & & $1.1 \times 10^{4}$ & & & "\\
        \hline \vspace{-0.2cm}
        & & & & & & & &\\
         \hline
        \textbf{CASE (3)} & & & & BACKGROUND      & & & & DUSTY & & & statistics\\
               & & & Wintertime & Summertime & Annual & & & Annual & & & running time\\
        \hline
        0.3 $\mu$m & & & $1.0 \times 10^{6}$ & $3.7 \times 10^{6}$ & $2.3 \times 10^{6}$ & & & $2.8 \times 10^{7}$ & & & Aug. 2001 $-$ Jan. 2010\\
        0.5 $\mu$m & & & $1.2 \times 10^{5}$ & $3.7 \times 10^{5}$ & $2.4 \times 10^{5}$ & & & $4.6 \times 10^{6}$ & & & "\\
        1.0 $\mu$m & & & $0.4 \times 10^{5}$ & $1.1 \times 10^{5}$ & $0.6 \times 10^{5}$ & & & $2.6 \times 10^{6}$ & & & "\\
        3.0 $\mu$m & & & $2.7 \times 10^{3}$ & $5.2 \times 10^{3}$ & $3.9 \times 10^{3}$ & & & $5.1 \times 10^{5}$ & & & Mar. 2007 $-$ Jan. 2010\\
        5.0 $\mu$m & & & $1.0 \times 10^{3}$ & $1.5 \times 10^{3}$ & $1.3 \times 10^{3}$ & & & $1.6 \times 10^{5}$ & & & Aug. 2001 $-$ Jan. 2010\\
       10.0 $\mu$m & & & $1.2 \times 10^{2}$ & $0.7 \times 10^{2}$ & $1.0 \times 10^{2}$ & & & $1.1 \times 10^{4}$ & & & Mar. 2007 $-$ Jan. 2010\\
        \hline
        \end{tabular}
          \label{cases}
        \end{center}
\end{table*}
   
\begin{figure*}
   \begin{center}
   \includegraphics[width=10cm]{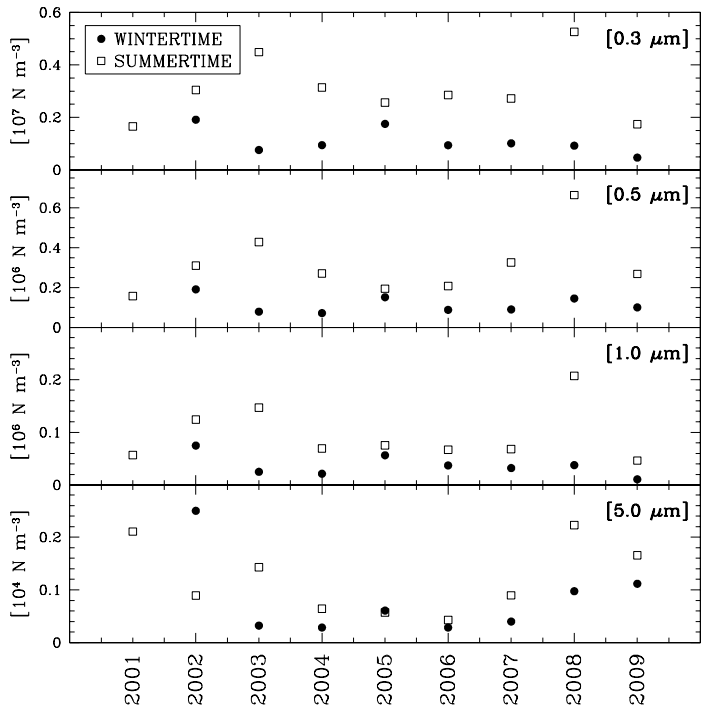}
    \caption{Seasonal dust background distribution at TNG during background dust conditions, in counts per cubic meter [N m$^{-3}$].}
   \label{seasonal}
   \end{center}
\end{figure*}

\section{Seasonal distribution of the dust in background and dusty conditions}\label{season_sec}

The raw data from the counters have been analyzed following the same method of Paper I. Because we cannot distinguish between the types of particles, but only their size,  we rejected data obtained when the measured relative humidity was greater than 85\%, that corresponds to the condensation point for water vapor particles.  As we will see later, this may introduce biases in the most humid months. Dust counts having a values of a few $\sigma$ above the median values have been classified as \textit{dust storms}. The background is evaluated using a $\kappa-\sigma$ clipping algorithm as described in Huber \cite{huber81};  Patat \cite{patat03};  and also used in Paper I.

Because particle counts  $N_{i}$ --where $i$ is the $i$th size bin-- have a Poissonian distribution, for each fixed period (year, semester, or month) we compute the median ($M$), and for each particle size we calculate the Median Absolute Deviation ($\Gamma$) defined as the median of the distribution $\left| N_{i} - M \right|$. For a Gaussian distribution, the ratio between the standard deviation and $\Gamma$ is equal to 1.48 (Huber\citealt{huber81}), so we set $\sigma = 1.48\Gamma$. We perform two iterations to compute the background rejecting in each iteration counts that exceed,

\begin{equation}
\left| N_{i} - M \right| > \kappa\sigma
\end{equation}

In the first iteration we set $\kappa=3$  and in the second we rejected measurements between  $-3\sigma$ and  $+2\sigma$ in order to finally get the background distribution, uncontaminated by dust peaks corresponding to dust storms. 

In order to investigate the presence of  seasonal variations of the background dust content,  we divided our database in wintertime (semester October-March) and summertime (semester April-September). Figure \ref{seasonal} shows the trend of the annual background distribution for the 8-years measured particles (0.3, 0.5, 1.0 and 5.0 $\mu$m).  We remark that in this plot we have removed dust storms using the $\kappa-\sigma$ clipping procedure described above, so these results correspond to strictly background dust conditions.  Figure~\ref{seasonal}  shows that the dust background is always larger in Summer than in Winter except in 2002, when the largest particles (5.0 $\mu$m) were significantly more abundant in winter than in summer.  Curiously, this anomaly is not present for the smaller particles. A further comparison with the TOMS (Total Ozone Mapping Spectrometer) data does not show the same result. It is important to mention that the TOMS aerosol index (AI) has as a major limitation the inability to detect dust occurring at or near the surface since the ground signal can overwhelm the dust signal (Herman et al.\citealt{herman97}; Gao\&Washington\citealt{gao10}). The 2002 dust recorded by our ground based dust sensor could be confined to the lowermost layers of the atmosphere, which might explain why it could not be seen by TOMS. Our guess is that the 2002 phenomenon corresponds to a local event not seen by the satellite (contamination from local recycled dust?).\\

Table \ref{cases} presents the dust background in counts per cubic meter [N m$^{-3}$] in background dust conditions
 distinguishing between wintertime, summertime, and an entire annual cycle, and also the dust content during dust storms in an entire annual cycle, for 3 different cases:
\begin{itemize}
   \item[\textbf{(1)}] using the Abacus TM301 database between August 2001 and December 2006 for particles of 0.3, 0.5, 1.0, and 5.0 $\mu$m sizes;\\
   \item[\textbf{(2)}] using the LasairII 310B database between March 2007 and January 2010 for particles of 0.3, 0.5, 1.0, 3.0, 5.0 and 10.0 $\mu$m sizes;\\
   \item[\textbf{(3)}] using both the Abacus TM301 and the LasairII 310B databases between August 2001 and January 2010 for particles of 0.3, 0.5, 1.0 and 5.0 $\mu$m sizes, and the LasairII 310B database for particles of 3.0 and 10.0 $\mu$m sizes.
\end{itemize}

\begin{figure}
    \begin{center}
         \includegraphics[width=8cm]{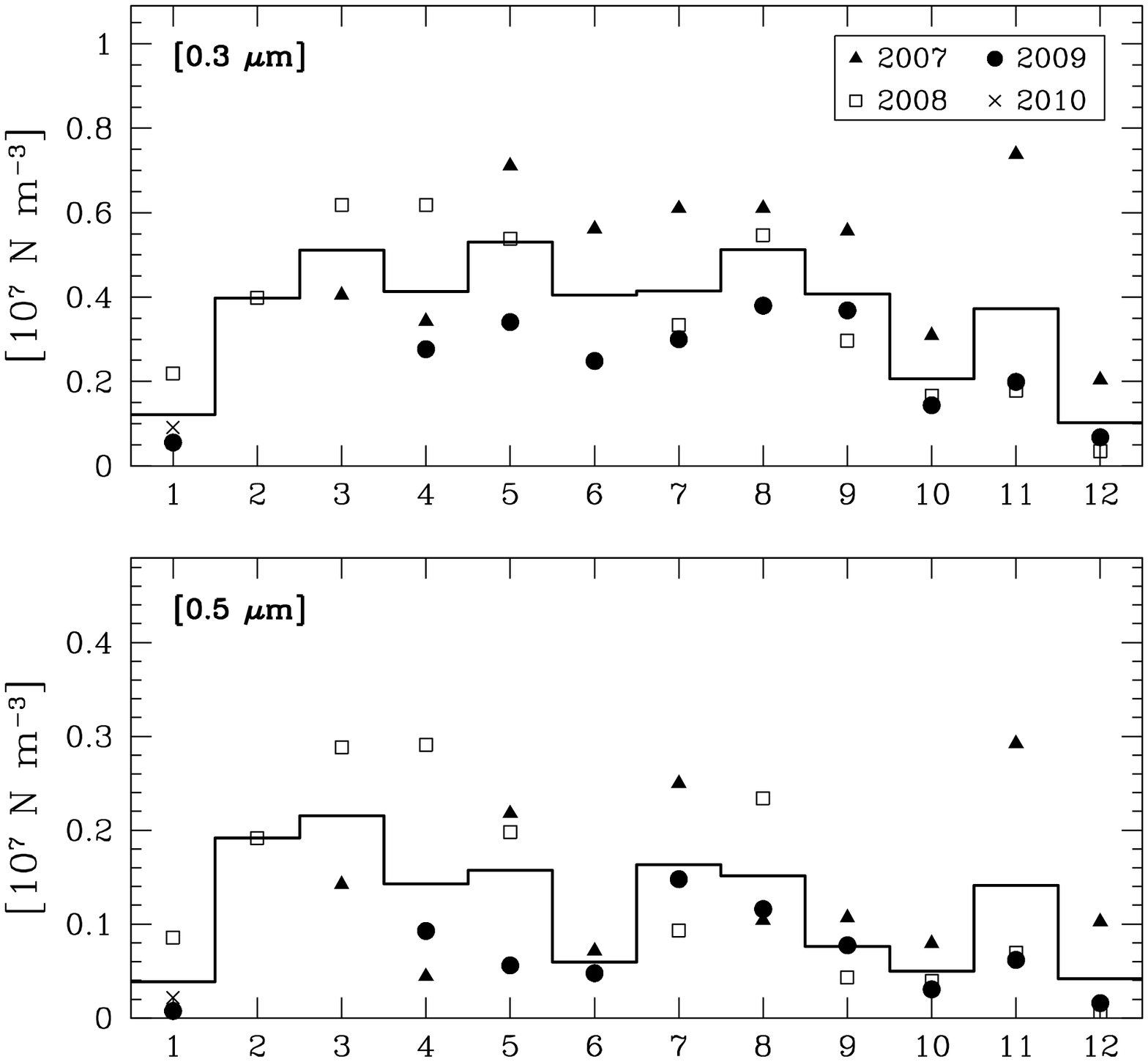}\vspace{-0.65cm}
         \includegraphics[width=8cm]{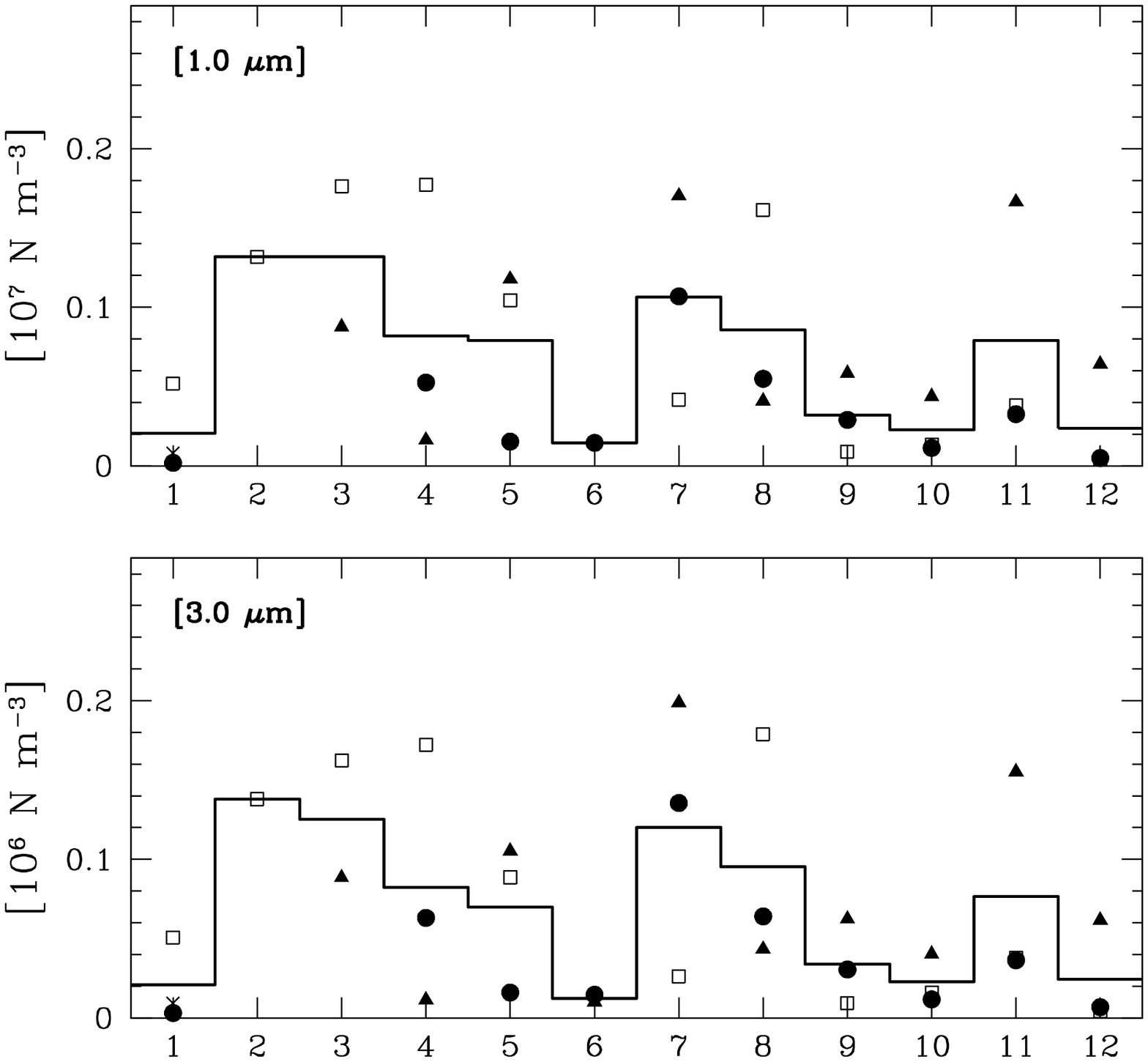}\vspace{-0.65cm}
         \includegraphics[width=8cm]{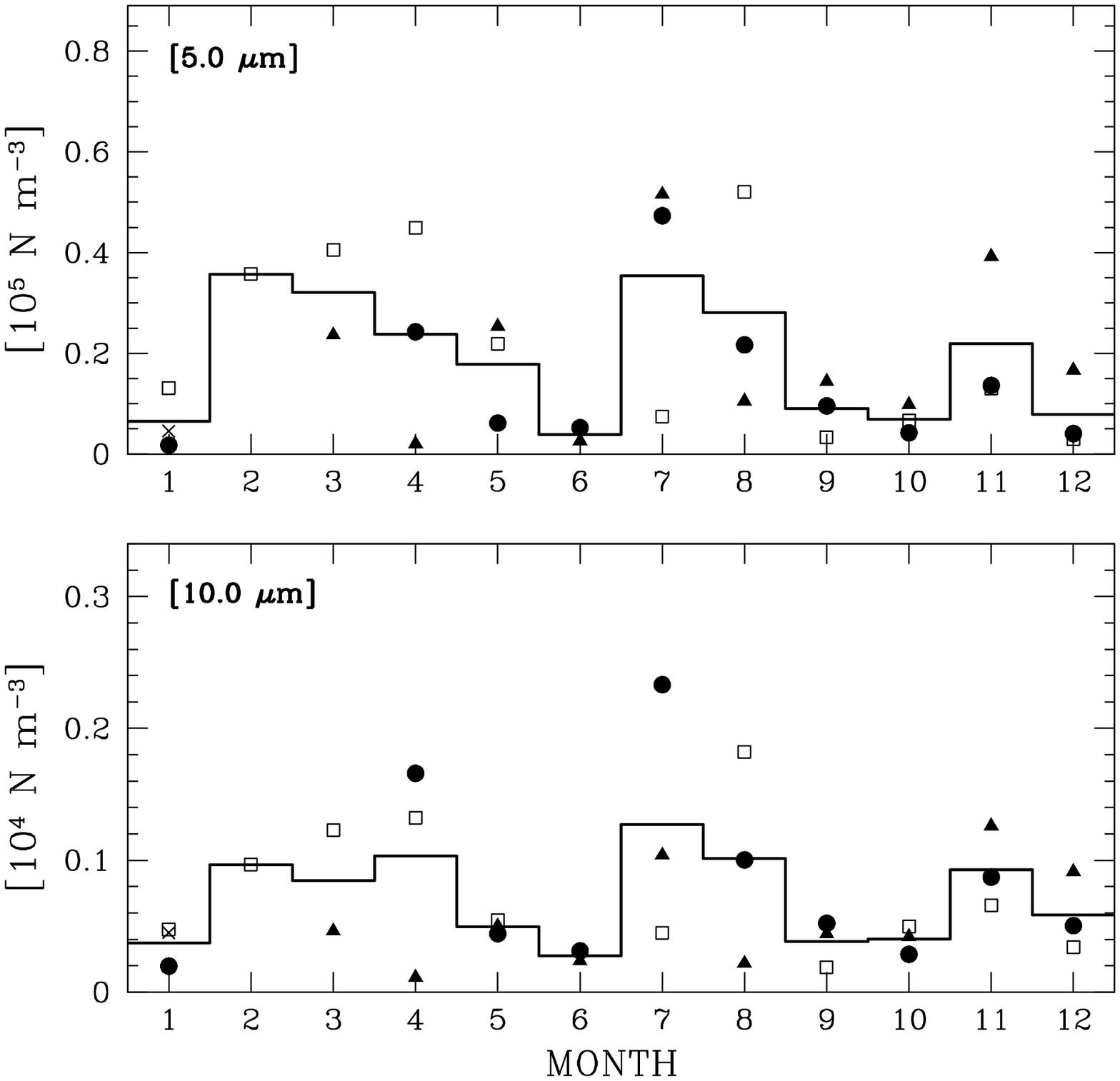}\vspace{-0.50cm}
    \caption{Seasonal distribution of the monthly dust at ORM in counts per cubic meter [N m$^{-3}$]. Different years are represented by different symbols. The solid line shows the monthly values averaging over all years.}
    \end{center}
    \label{monthly}
\end{figure}

Figure \ref{seasonal} and Table \ref{cases} show that the prevailing aerosol conditions at ORM have not changed significantly during the 8 years of monitoring, with summertime being  dustier  than wintertime, with the exception of the 10.0 $\mu$m particles that show the opposite behavior. 

We now consider the monthly median counts without excluding the dust storms. In Paper~I we found that between 2001 and 2006 the monthly distribution of aerosols is characterized by a significant increase in the counts during February-April and July-August of each year. Figure 2 shows the median monthly distribution of particles from the LasairII 310B data that confirms the trends found in Paper~I, with the exception of November 2007 that was  clearly a very dusty month.  June, September, October, and December-January appear to be the cleanest months, but we remark that December and January are also  the most humid months, so these results may be biased because the dust counters are not reliable (relative humidity greater than 85\%).

\begin{figure}
   \begin{center}
   \includegraphics[width=8cm]{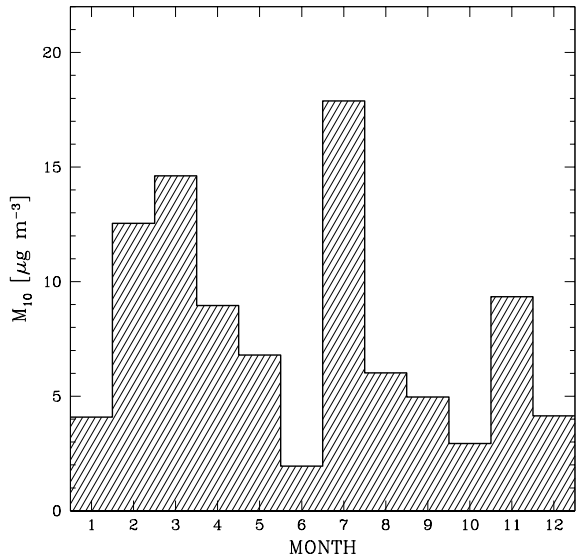}	
   \caption{Monthly distribution of dust density at ORM computed adding together all particle sizes for Case \textbf{(2)}. A constant density of 2.5 g cm$^{-3}$ was assumed for all particle sizes.}
   \end{center}
   \label{density}
\end{figure}

Figure \ref{density} shows the monthly distribution of overall dust density $M_{10}$ (in micro grams per cubic meter) computed adding together all particle sizes for Case \textbf{(2)} and assuming that all particles have the same density of 2.5 g cm$^{-3}$, which is typical of silicates and quartz aggregates (Suh\citealt{suh99}) that are the main components of Saharan dust (Murdin\citealt{murdin85}).  In this plot July emerges as the dustiest month, while somewhat surprisingly June is the cleanest. As already discussed, the Spring is dustier than the Summer and dusty conditions are also frequent in the Winter.

\section{Thermal background in the NIR}\label{emissivity}

A dust particle of radius $a$ is heated by an ambient gas of molecules at temperature $T_m$ at a rate that depends
on the number density of molecules and the fraction of energy $E$ that is deposited in the dust grain by the
impinging particles (Dwek\citealt{dwek86}). The dust particle radiates in the infrared the energy acquired in the collision at a rate that depends on
the grain temperature $T_d$ as,

\begin{equation}
E_* = 4 \pi a^2 \int_{\Delta\lambda}\pi B_\lambda (T_d) Q_\lambda (a) d\lambda
\end{equation}
where $B_\lambda (T_d)$ is the Planck function for temperature $T_d$ and $Q_\lambda (a)$ is the grain emissivity at wavelength $\lambda$ that is usually calculated using Mie theory (Dewk\citealt{dwek86}).  In this  Section we estimate the thermal background emission of atmospheric dust in the  NIR spectral bands, for which we will use the 2MASS photometric system reproduced in Table~\ref{2MASS}.
 
\begin{table}
     \begin{center}
     \caption[]{Absolute calibration of the 2MASS photometric system taken from  Cohen et al. \cite{cohen03}.}
     \label{2MASS}
     \begin{tabular}{c | c c | c}
\hline
 Filter & $\lambda_{eff}$ & $\Delta \lambda$ & $F_{0,\lambda}$\\
        &  [$\mu$m] & [$\mu$m]         & [W m$^{-2}$ $\mu$m$^{-1}$]\\
 \hline
 $J$       & 1.235     & 0.162  & $3.129 \times 10^{-9}$\\
 $H$      & 1.662     & 0.251  & $1.133 \times 10^{-9}$\\
 $K_S$ & 2.159     & 0.262  & $4.283 \times 10^{-10}$\\
\hline
    \end{tabular}
    \end{center}
\end{table}

We can use our measured dust densities to calculate the total dust optical depth $\tau_\lambda$  as the sum of the contribution of the optical depth of each particle size,

\begin{equation}
\tau_\lambda = \sum_{a}\tau_\lambda (a) = \sum_{a} N(a)\sigma_\lambda (a)
\end{equation}
where $N(a)$ is the  column density of particles of radius $a$, and $\sigma_\lambda (a)$ is the wavelength dependent absorption cross section from the Mie theory for particles of radius $a$.

Satellite measurements indicate that the dust density over the Canary Islands  is approximately constant at altitudes between 2500 m and 5000 m above sea level, and drops to virtually zero above 5000 m (Smirnov et al.\citealt{smirnov98}; Hsu et al.\citealt{hsu99}; Alpert et al.\citealt{alpert04}; Paper~I).   Thus, the column densities are simply the volume densities  multiplied by 2500 m. The resulting optical depths are calculated using the dust content in background and dusty conditions obtained for an entire annual cycle  and shown in Case \textbf{(3)} of Table \ref{cases}. Results  are given in Table \ref{tau} and are in very good agreement with those in the visible from Smirnov et al. \cite{smirnov98} calculated above 2356 meters above sea level in Tenerife Canarian Island (see Table 2 in the mentioned paper). The extrapolation to NIR wavelengths still remains in good agreement with our results.

\begin{table}
     \begin{center}
     \caption[]{Aerosol optical depth calculated for the 2MASS  filters.}
     \begin{tabular}{c | c c }
\hline
 Filter & background  & dusty \\
 \hline
 $J$    & 0.006 & 0.296\\
 $H$    & 0.005 & 0.280\\
 $K_S$  & 0.004 & 0.270\\
\hline
    \end{tabular}
         \label{tau}
    \end{center}
\end{table}

The cooling times for dust particles of the sizes and temperatures typical of the Saharan dust above ORM are very short (Kaiser\citealt{kaiser70}; Draine\citealt{draine81}; Dewk\citealt{dwek86}), so it is reasonable to assume that the dust is locally in thermal equilibrium with the surrounding air, while clearly the dust is optically thin even under the dustiest conditions. Thus, the thermal radiation emitted by the dust particles can be calculated as,

 \begin{equation}
I_\lambda = \tau_\lambda B_\lambda (T_d) =  \frac{2c^{2}h}{\lambda^{5}}\frac{1-e^{-\tau_{\lambda}}}{\exp{(\frac{hc}{k_{B} \lambda T_d})}-1}
\end{equation}
where $B_\lambda (T_d)$ from Tokunaga \cite{tokunaga00} is expressed in [W m$^{-2}$ $\mu$m$^{-1}$ sr$^{-1}$], $c$ is the light speed in vacuum,
$h$ is the Planck's constant, and $k_B$ is the Boltzmann constant.
 
We must calculate the Planck function considering that $T_d$ decreases with altitude ($h$) from the median value at the ground (about 282 K, see Lombardi et al.\citealt{lombardi06}) with a wet vertical adiabatic lapse rate of $-0.006$ K m$^{-1}$ (Kittel\&Kroemer\citealt{kittel80}).  Thus, the thermal spectrum of the dust is given by the expression,

\begin{equation}\label{Flux4}
F_\lambda  = \sum_{a}\tau_\lambda(a) \int_{0}^{2500}  B_\lambda (T_d(h)) dh
\end{equation}
where the sum is over all particle sizes, $a$. Using the absolute calibration of the 2MASS photometric system presented  in Table \ref{2MASS} we obtain the aerosols sky brightness in the NIR in background and dusty days as,

\begin{equation}\label{magmag}
m_\lambda = -2.5 ~ \log{\frac{F_\lambda}{F_{0,\lambda}}}
\end{equation}
The results are shown in Table \ref{NIRemissivity} that clearly show that the sky brightness due to dust in the atmosphere is significant only during dusty conditions, and only in the $K_S$ band. From the TNG data archive we know that the sky background in $J$ is between 15.0 and 16.0 mag arcsec$^{-2}$, in $H$ is between 13.4 and 14.7 mag arcsec$^{-2}$, while in $K_S$ it varies from 12.5 and 13.0 mag arcsec$^{-2}$. We conclude that at ORM the contribution of the dust to the sky background in the NIR is mostly negligible in both background and dusty conditions, with a maximum contribution of about 8-10\% in the $K_{S}$ band in the dusty days.

\begin{table}
     \begin{center}
     \caption[]{Aerosols background NIR emissions in [mag arcsec$^{-2}$] in background and dusty days.}
  
     \begin{tabular}{c | c c}
     \hline
 Filter & background conditions & dusty days\\
\hline
 $J$   & 39.2 & 35.2 \\
 $H$   & 27.8 & 23.6 \\
 $K_S$ & 20.3 & 15.8 \\
\hline
    \end{tabular}
       \label{NIRemissivity}
    \end{center}
\end{table}

\begin{table}
     \begin{center}
     \caption[]{Aerosols optical depth and NIR emissions in [mag arcsec$^{-2}$] for the major event occurred between 25 and 30 July 2007.}
  
     \begin{tabular}{c | c c}
     \hline
 Filter & $\tau_{\lambda}$ & emission\\
\hline
 $J$   & 0.388 & 34.5 \\
 $H$   & 0.372 & 23.0 \\
 $K_S$ & 0.356 & 15.2 \\
\hline
    \end{tabular}
       \label{major}
    \end{center}
\end{table}
Given that our results were computed over a long period, the averaging process could have smoothed larger events, so the full radiative impact of these events may go undetected. As an example, we have calculated the optical depth (and thus the resulting thermal emissivity) for a major event occurred between 25 and 30 July 2007. Results are reported in Table \ref{major} and confirm that still during a major event the contribution of the dust to the sky background in the NIR is not critical.

We can conclude that the obtained thermal emission is the contribution of our in situ measured particles and results could be intended 
as a lower limit.

\section{Conclusions}
Using a laser diode particle counter near the TNG at ORM  we have measured the densities of airborne aerosols of 0.3, 0.5, 1.0, 3.0, 5.0 and 10.0 $\mu$m size. The seasonal 
trends of the particle content in the atmosphere have not changed significantly between 2001 and 2010.  The monthly distribution of aerosols is characterized by an increase during February-April 
and July-August of each year: the Spring is dustier than the Summer, but dusty conditions may also occur in Winter.

Using the Mie theory we calculated the dust absorption cross section and thus estimated the thermal emission in the 2MASS NIR spectral bands. Assuming that the dust particles are locally in thermal equilibrium with the surrounding air, and that the air temperature decreases with altitude with the wet vertical adiabatic lapse ($-0.006$ K m$^{-1}$), we found that the contribution of dust emission to the total sky background in the NIR is negligible comparable to the airglow component during both background and dusty conditions, with a maximum contribution of about 8-10\% in the $K_{S}$ band in the dusty days.

\section*{Acknowledgments}
\vspace{-0.15cm}
The authors aknowledge the anonymous reviewer for the very helpful comments on the paper.

\bsp

\label{lastpage}


\begin{thebibliography}{99}\footnotesize
\vspace{-0.15cm}
\bibitem[2004]{alpert04} Alpert, P., Kishcha, P., Shtivelman, A., Krichak, S. O., \& Joseph, J. H. 2004, Atmos. Res., 70, 109
\bibitem[2003]{cohen03} Cohen, M., Wheaton, W. A., \& Megeath, S. T. 2003, AJ, 126, 1090
\bibitem[2009]{cuevas09} Cuevas E., \& Baldasano J. M. 2009, \textit{Report on the Incidence of African dust intrusions at the Astronomical Observatories of teh Canary Islands: characterization and temporal analysis}, Report released by the Iza\~na Atmospheric Research Center (Meteorological State Agency of Spain; AEMET) and the Earth Sciences Department (Barcelona Supercomputing Center -- Centro Nacional de Supercomputaci\'on; BSC-CNS) under request of the Instituto de Astrofisica de Canarias (IAC)
\bibitem[1986]{dwek86} Dwek, E. 1986, ApJ, 302, 363
\bibitem[1981]{draine81} Draine, B. T. 1981, ApJ, 245, 880
\bibitem[2010]{gao10} Gao, H., \& Washington, R. 2010, Climate Dynamics, 35(2-3), 511
\bibitem[2003]{gelado03} Gelado, M. D., Dorta, P. J., Torres, M.E., Hern\'andez, J. J., Collado, C., Siruela, V. F., Cardona, P., \& Rodr\'iguez, M. J. 2003, \textit{Caracterizaci\'on del aerosol Sahariano en Gran Canaria}, Encuentro sobre Meteorologia y Atm\'osfera de Canarias, http://acceda.ulpgc.es/bitstream/10553/1531/1/5391.pdf
\bibitem[2004]{ghedina04} Ghedina, A., Pedani, M., Guerra, J. C., Zitelli, V., \& Porceddu, I. 2004, SPIE, 5489, 227
\bibitem[1998]{guerrero98} Guerrero, M. A., Garc\'ia-L\'opez, R. J., Corradi, R. L. M., Jim\'enez, A., Fuensalida, J. J., et al., 1998, New Astron. Rev., 42, 529
\bibitem[1998]{jimenez98} Jim\'enez, A., Gonzalez Jorge, H., \& Rabello-Soares, M. C. 1998, A\&AS, 129, 413
\bibitem[1997]{herman97} Herman, J., Bhartia, P., Torres, O., Hsu, C., Seftor, C., \& Celarier, E. 1997, J. Geophys. Res., 102(D14), 16911
\bibitem[1981]{huber81} Huber, P. J. 1981, Robust Statistics, ed. J. Wiley \& Sons, NY
\bibitem[1999]{hsu99} Hsu, N. C., Herman, J. R., Torres, O., Holben, B. N., Tanre, D., \& Eck, T. F. 1999, J. Geophys. Res., 104, 6269
\bibitem[1970]{kaiser70} Kaiser, C. B. 1970, ApJ, 159, 77
\bibitem[1980]{kittel80} Kittel, C., \& Kroemer, H. 1980, Thermal Physics, ed. W. H. Freeman Company (2nd ed.)
\bibitem[2006]{lombardi06} Lombardi, G., Zitelli, V., Ortolani, S., \& Pedani, M. 2006, PASP, 118, 1198
\bibitem[2008]{lombardi08} Lombardi, G., Zitelli, V., Ortolani, S., Pedani, M., \& Ghedina, A. 2008, A\&A, 483, 651 (Paper I)
\bibitem[1985]{murdin85} Murdin, P. 1985, Vistas Astron., 28, 449
\bibitem[2003]{patat03} Patat, F. 2003, A\&A, 401, 797
\bibitem[2002]{porceddu02} Porceddu, I., Zitelli, V., Buffa, F., \& Ghedina, A. 2002, SPIE, 4844, 358
\bibitem[2002]{smirnov98} Smirnov, A., Holben, B. N., Slutsker, I. , Welton, E. J., \& Formenti, P. 1998, J. Geophys. Res., 103(D21), 28, 079-28, 092
\bibitem[1999]{suh99} Suh, K. W. 1999, MNRAS, 304, 389 
\bibitem[1987]{stickland87} Stickland, D. J., Lloyd, C., Pike., C. D., \& Walker, E. N. 1987, The Observatory, 107, 74
\bibitem[2000]{tokunaga00} Tokunaga, A. T. 2000, in Allen's Astrophysical Quantities, ed. A. N. Cox (4th ed.; New York: AIP) 

\end{thebibliography}
\end{document}